\documentclass[sigconf, language=french,
language=german, language=spanish, language=english]{acmart}

\AtBeginDocument{%
  }

\setcopyright{acmlicensed}
\copyrightyear{2025}
\acmYear{2025}

\acmConference[XX]{Make sure to enter the correct
  conference title from your rights confirmation emai}{XX}

\begin{document}

\title{Enhancing IoT Network Security through Adaptive Curriculum Learning and XAI}

\author{Sathwik Narkedimilli}
\email{21bcs103@iiitdwd.ac.in}
\affiliation{%
  \institution{Department of Computer Science, Indian Institute of Information Technology (IIIT) Dharwad}
  \city{Dharwad}
  \country{India}
}

\author{Sujith Makam}
\email{21bcs061@iiitdwd.ac.in}
\affiliation{%
  \institution{Department of Computer Science, Indian Institute of Information Technology (IIIT) Dharwad}
  \city{Dharwad}
  \country{India}
}

\author{Amballa Venkata Sriram}
\email{21bcs008@iiitdwd.ac.in}
\affiliation{%
  \institution{Department of Computer Science, Indian Institute of Information Technology (IIIT) Dharwad}
  \city{Dharwad}
  \country{India}
}

\author{Sai Prashanth Mallellu}
\email{saiprashanth08@ieee.org}
\affiliation{%
  \institution{Department of Computer Science \& Engineering, Vardhaman College of Engineering}
  \city{Hyderabad}
  \country{India}
}

\author{MSVPJ Sathvik}
\email{msvpjsathvik@gmail.com}
\affiliation{%
  \institution{Department of Computer Science, Indian Institute of Information Technology (IIIT) Dharwad}
  \city{Dharwad}
  \country{India}
}

\author{Ranga Rao Venkatesha Prasad}
\email{R.R.VenkateshaPrasad@tudelft.nl}
\affiliation{%
  \institution{TU Delft}
  \country{Netherlands}
}

\renewcommand{\shortauthors}{Sathwik et al.}

\begin{abstract}
To address the critical need for secure IoT networks, this study presents a scalable and lightweight curriculum learning framework enhanced with Explainable AI (XAI) techniques, including LIME, to ensure transparency and adaptability. The proposed model employs novel neural network architecture utilized at every stage of Curriculum Learning to efficiently capture and focus on both short- and long-term temporal dependencies, improve learning stability, and enhance accuracy while remaining lightweight and robust against noise in sequential IoT data. Robustness is achieved through staged learning, where the model iteratively refines itself by removing low-relevance features and optimizing performance. The workflow includes edge-optimized quantization and pruning to ensure portability that could easily be deployed in the edge-IoT devices. An ensemble model incorporating Random Forest, XGBoost, and the staged learning base further enhances generalization. Experimental results demonstrate 98\% accuracy on CIC-IoV-2024 and CIC-APT-IIoT-2024 datasets and 97\% on EDGE-IIoT, establishing this framework as a robust, transparent, and high-performance solution for IoT network security.
\end{abstract}


\maketitle

\section{Introduction}

The rapid growth of IoT networks~\cite{madakam2015internet} ~\cite{gurunath2018overview}~\cite{hamza2020iot} has introduced significant security challenges due to their distributed nature, resource constraints, and vulnerability to a wide range of cyber threats, including data breaches, denial-of-service attacks, and malicious intrusions. Traditional intrusion detection systems (IDS) struggle with scalability, adaptability, and transparency, making them less effective in handling the dynamic and diverse nature of IoT data~\cite{chen2024network}~\cite{chaabouni2019network}. Additionally, the lack of explainability and transparency in many models' predictions, limits their usability in critical applications, where trust and interpretability are essential~\cite{wheelus2020iot}.

In response to such problems, this research develops a new lightweight and scalable curriculum learning system that progressively refines model performance. The framework begins with simpler attack scenarios, gradually advancing to more complex threats, thereby enhancing the model's robustness and flexibility and it is enhanced by Explainable AI (XAI) methods like LIME. The model combines deep neural network structures (such as GRU, LSTM, and self-attention) to capture IoT time-dependent dependences at momentary and long-term levels. Using staged learning, adaptive feature masking, and edge-optimized quantization, the model incrementally becomes stronger and more efficient. The ensemble model also gives more generalization as it blends the power of the staged learning model and classifiers such as Random Forest or XGBoost.

The motivation for this research is that the IoT devices are getting more and more mainstream, in crucial industries like healthcare, smart cities, and industrial automation, and security is crucial. In addition, the need for an interpretable and adaptable IDS solution for resource-dense environments and attack scenarios is just one of the requisites to combining curriculum knowledge with XAI. By making sure security products are accurate, also transparent, portable, and trustworthy.

The framework can be broadly used to secure IoT ecosystems like home automation, industrial IoT (IIoT), health surveillance, and self-driving cars~\cite{wang2019software}~\cite{pamarthi2022literature}~\cite{hassija2019survey}. XAI can be added to make the predictions clear and understandable by stakeholders so that they can better predict and mitigate security risks. The thin architecture ensures deployment in edge devices for scalability and real-time threat intelligence in IoT landscapes of various sizes. The study's implications go beyond improving trust in AI-based security systems and providing an example of learning in IoT networks.

This research contributes to the concept of adaptive staged learning of the key pillars of curriculum learning as it builds the model over time from low to high-risk attack scenarios. Key improvements are (1) adaptive masking feature selector with dynamic features selection, (2) GRU, LSTM, and attention functions to capture a wide variety of temporal patterns, (3) edge-optimized quantization and pruning for lightweight implementation, (4) model transparency and trust enhancement by XAI and (5) ensemble model with staged learning and classifiers to generalize. These features give a highly accurate, scalable, and readable solution for IoT network security with high precision and elasticity on a range of datasets.

The study flows as follows: Section~\ref{sec1} introduces preliminaries, Section~\ref{sec2} reviews related work, Section~\ref{sec3} details the proposed model, Section~\ref{sec4} evaluates and analyzes its performance, Section~\ref{sec5} concludes the findings, and Section~\ref{sec6} discusses future research directions.

\section{Preliminaries} \label{sec1}

\subsection{Curriculum Learning}

Curriculum learning~\cite{soviany2022curriculum} ~\cite{bengio2009curriculum} is based on human education is a machine-learning approach in which models are trained in stages, first with simpler tasks and then with higher-level tasks. This staged approach improves model stability, generalization, and robustness by teaching the system basic patterns before it tackles complex cases~\cite{hacohen2019power}~\cite{polu2022formal}. Curriculum learning can especially help network security because it can adjust models to evolving IoT network attack scenarios. : Training on low-risk attacks first, then moving to high-risk, complex threats, improves detection accuracy and scalability. This approach is especially appropriate for resource-limited edge environments due to feature selection and learning efficiency which is a good technique to protect IoT ecosystems from sophisticated cyber attacks~\cite{zhou2021curriculum}.

\subsection{Explainable Artificial Intelligence (XAI)}

Explainable AI (XAI)~\cite{zhang2022explainable} ~\cite{charmet2022explainable}: XAI attempts to enable machine learning models to be comprehensible and understandable, and able to explain how they make decisions. LIME (Local Interpretable Model-agnostic Explanations)~\cite{nagaraj2022prediction}~\cite{dieber2020model} is one popular XAI method to give an account of predictions by replacing hard-to-apprehend models with locally linear models on a particular example. LIME is very important for network security because LIME helps detect features that are important for model choices like a particular pattern of network traffic pointing toward an attack. In human-readable terms, LIME makes automated intrusion detection more trustable, allows cybersecurity professionals to confirm model behavior, and enables rapid response to new threats in IoT networks. That interoperability is important when it comes to the security of sensitive applications and establishing trust in AI-based security.

\section{Literature Review} \label{sec2}

The following studies explore advancements in IoT network security, the evolving threat landscape, and commonly employed methods to address these challenges.

The research study "A Survey on Cybersecurity in IoT"~\cite{dritsas2025survey} by Elias et al.: describes the proliferation of IoT devices and brings security issues because they are distributed, and heterogeneous and the threat landscape is growing. Security – IoT systems are normally not very secure, which leaves them at risk of hacking, DDoS attacks, and illegal monitoring. Computing power is scarce on devices that can’t be used for effective security measures, and devices are tied to wireless communications and vulnerable to eavesdropping and man-in-the-middle attacks. Such problems will require new techniques like light cryptography, blockchain-based decentralized security, and AI-powered anomaly detection. However, there are some constraints, such as scaling, high computation, and integration overhead. Such limitations demand responsive, economical approaches to protect IoT ecosystems as attack surfaces evolve. The above study describes the need for a robust method yet adaptable to the modern threat landscape in the IoT network scenarios,

The research article "Defence and Security Mechanisms in the Internet of Things: A Review"~\cite{szymoniak2025defense} conducted by Sabina et al. describes the ever-increasing avalanche of IoT attacks are different from the other due to the proliferation of smart devices and their penetration into the healthcare, industrial and smart city sectors. These are the most susceptible devices to a large variety of cyber threats like DoS attacks, impersonation, session key leaks, and insider threats and are also vulnerable to limited computing power and multiple communication protocols. Our approach to these problems leverages secure cryptographic protocols (e.g., ECC, blockchain authentication) with advanced anomaly detection methods that make use of AI like gradient boosting, LSTMs, and CNNs to detect and mitigate abnormal activity. Nevertheless, limitations still exist such as the high computational cost of AI solutions, scaling problems in resource-limited systems, and difficulty of interoperability between diverse IoT ecosystems. These limitations require thin, flexible, and adaptive security architectures to protect IoT devices from an ever-shifting threat environment.

The following studies outline frameworks for network security, leveraging novel approaches such as Explainable AI (XAI) and ensemble techniques.

Muhammet et al.~\cite{yagiz2025lens} proposed the LENS-XAI research strategy, which integrates an encapsulated, transparent intrusion detection algorithm via knowledge distillation and variational autoencoders (VAEs). The research normalizes features and encodes categorical variables, and then passes them into a VAE for latent representation learning. These are reduced to a student model, which reduces the number of computations needed in resource-limited situations. The system is designed with explanations in mind by variable attribution methods and gives a transparent view of anomaly detection. Even with such high detection accuracy and effectiveness, the disadvantages are difficulties in rare attack detection and dependency on benchmark data, which might compromise generalizability to different scenarios in the real world.

The algorithm used in the research article "Machine Learning and Metaheuristic Optimization Algorithms for Feature Selection and Botnet Attack Detection"~\cite{maazalahi2025machine} by Mahdieh et al.: a combination of Machine Learning, ensembled learning, and Metaheuristic Optimization (ML) based hybrid IDS. It has 3 steps: Preprocessing (normalization and elimination of outlier by K-NN), feature selection (synthesis of SailFish Optimizer and Whale Optimization Algorithm), and attack detection (synthesis of PSO-K-means). This algorithm is tested on BOT-IOT and UNSW-NB15 datasets with high accuracy and performance in botnet attack detection. However, constraints are the dependence of the first population diversity on optimization algorithms and the computation challenges of working with high-dimensional datasets.

In the research paper "Diwall: A Lightweight Host Intrusion Detection System Against Jamming and Packet Injection Attacks,"~\cite{el2025diwall} a hardware HIDS is used to protect IoT end-devices. Methodology: RISC-V processor, LoRaWAN protocol, monitoring microarchitectural activities, and network data (RSSI) data. The system also uses a decision tree and exponentially weighted moving average to detect memory corruption, packet injection, and jamming attacks. However the study’s flaws are limited by the fact that its lightweight solution is not scalable to larger IoT protocol stacks, it might be hardware-limited, and it’s hard to extend the method to more advanced attacks or devices with increased computational requirements.

The studies "Malware Detection of IoT Networks Using Machine Learning: An Experimental Study with Edge-IIoT Dataset"~\cite{hamza2023malware} and "DDoS Attack Detection in Edge-IIoT Network Using Ensemble Learning"~\cite{laiq2024ddos} explore machine learning approaches for securing IoT networks. Both employ comprehensive preprocessing techniques, including feature scaling, label encoding, and handling data imbalance using SMOTE. The first study evaluates supervised algorithms like KNN, DTC, LR, SVM, and RFC, with RFC achieving a 94.1\% accuracy, demonstrating its effectiveness in malware detection. The second study leverages ensemble learning methods such as bagging, boosting, and stacking for DDoS attack detection, achieving high accuracy metrics. However, limitations include reliance on a single dataset, raising concerns about generalizability, and the need for additional evaluations on computational efficiency and scalability across diverse IoT environments.

The studies "Secure IIoT Networks with Hybrid CNN-GRU Model Using Edge-IIoTset"~\cite{saadouni2023secure} and "Binary and Multiclass Classification of Attacks in Edge IIoT Networks"~\cite{aslam2024binary} focus on enhancing intrusion detection in Industrial IoT (IIoT) networks using the Edge-IIoT dataset. The first study employs a hybrid deep learning model combining CNN for spatial feature extraction and GRU for handling sequential dependencies, achieving 98.70\% accuracy for multi-class classification with low false positive rates. The second study utilizes machine learning (ML) algorithms, including Random Forest and Gradient Boosting, alongside deep learning (DL) models like CNN, RNN, and ANN, achieving over 95\% accuracy for binary and up to 94\% for multi-class classification. Both studies emphasize comprehensive preprocessing techniques, including normalization and SMOTE for imbalance handling, but share limitations such as reliance on a single dataset and the need for broader evaluations to ensure generalizability and robustness in diverse IoT environments.

Although it has high accuracy in research studies, it isn’t scaleable and cost-effective to be deployed on limited resources edge devices. Also, these models aren’t transparent in their predictions — this is why you need XAI for decisions to help you understand and trust important IoT applications. Curriculum learning can combat these issues by training models step by step, beginning with low-risk attack scenarios and gradually working up to high-risk attacks. Not only does this optimize scalability by automating learning, but it also improves robustness and flexibility, which is ideal for edge-based environments with limited resources.

\section{Proposed Model} \label{sec3}

\subsection{Data-Preprocessing and Feature Engineering}

We have performed the pre-processing steps by initially adding missing values to zeros, erasing duplicates, converting data types, and discarding outliers for data consistency. Feature scaling is used by the StandardScaler algorithm to uniformize feature values to make the model efficient. Data reduction methods such as Linear Discriminant Analysis (LDA) reduce feature dimension where the number of elements is a function of explained variance to save the most important information. This processed data gets split up into training and testing data by an 80-20 split while maintaining class distributions with stratified sampling. Also, another validation set is developed for model performances during training.  

Feature engineering converts raw data into a more tractable model format for machine learning. This starts with feature selection – LDA plots or Random Forest feature importance plots for discovering and storing relevant predictors. The curriculum learning algorithm is applied to 50 features in the Edge-IIoT dataset. Dimensionality reduction is then further compressed using log scaling and binning to make the data more manageable. Features that have low predictive value are deleted or folded and interaction features are introduced to define inter-variable relationships. The resulting dataset has features (X) and target variables (y) that are explicitly tagged for supervised learning. This feature engineering and preprocessing pipeline provides data that is clean, optimized, and extremely effective in the curriculum learning models.

In this study, we evaluated the proposed framework on the Edge-IIoT, CIC-APT-IIoT-2024, and CIC-IoV-2024 datasets. Below, we outline the bifurcation of each dataset into various stages for curriculum learning, enabling the implementation of the proposed framework effectively.

In this study, the proposed algorithm was tested and implemented on the Edge-IIoT dataset, which was divided into four stages for curriculum training. Stage 1 focuses on training with normal data to establish a baseline model. Stage 2 incorporates simple attack data, including OS Fingerprinting, Port Scanning, and Vulnerability Scanner attacks. Stage 3 progresses to medium-level attacks, such as XSS, SQL Injection, Password, and Uploading attacks. Finally, Stage 4 addresses complex attack scenarios, including Backdoor, DDoS, MITM, and Ransomware, ensuring a comprehensive learning process across varying attack complexities.

The framework was further evaluated on the CIC-APT-IIoT-2024 dataset, where training was performed on both normal and attack data as two stages.

For the CIC-IoV-2024 dataset, we ran a 3-step training process for the attack cases. Training on normal data (Stage 1), Securing attacks with GAS, RPM, SPEED, STEERING WHEEL (Stage 2)). Stage 3: The model was trained on DoS attacks with the test data. Having such a structure, the model could be trained and updated to ever more IoV attack scenarios proving it was scalable and efficient.

\subsection{Workflow of the Proposed Model}

\begin{figure*}
    \centering
    \fbox{\includegraphics[width=1\linewidth]{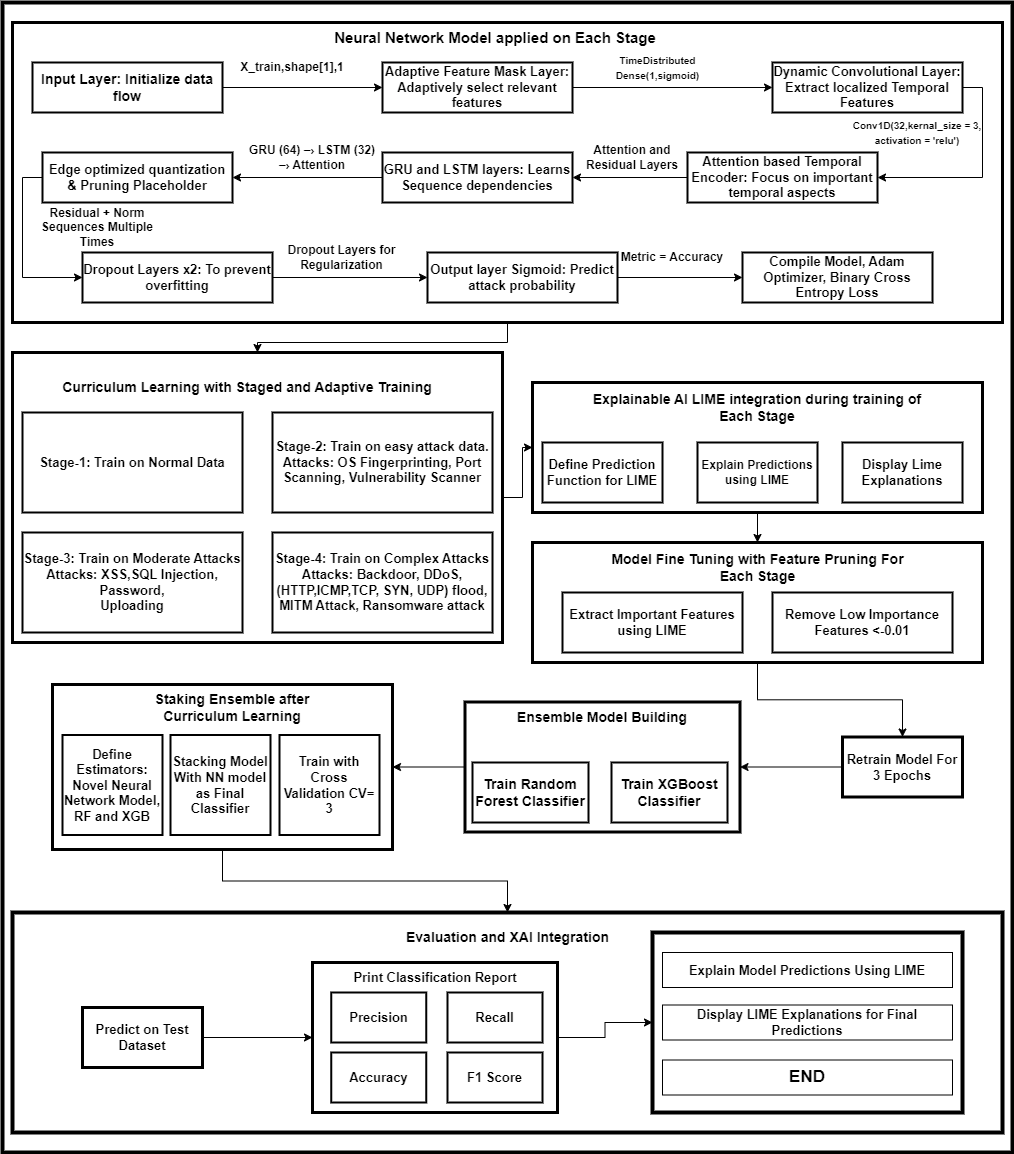}}
    \caption{The workflow of the Proposed Framework} \label{figu1}
\end{figure*}

Please refer to Fig. ~\ref{figu1} for an illustration of the workflow of the proposed Curriculum Learning algorithm. The proposed curriculum learning framework is designed for binary classification, where normal data is labeled as '0' and attack data is labeled as '1' and the proposed framework starts by utilizing the neural network is composed of advanced GRU, LSTM, Attention, and ResNet-like residual connections to deal with the sequential IoT data. It starts with the adaptive feature mask layer, which is a dense time-distributed layer that randomly ranks input features in order of significance. Then comes a dynamic convolutional layer which takes local time features to allow the model to recognize short-term dependencies that help identify attacks. The attention-based temporal encoder is used to sense long-term dependencies and maintain temporal relationships to make precise, context-dependent predictions for intricate IoT problems. 

And for even more learning we include residual connections and normalization layers after each block to smooth gradients and train in the stability mode. There are 3 GRU and LSTM layers included in the model; GRU layers can learn short-term dependencies using fewer parameters and LSTM layers can learn long-term dependencies. Self-attention functions post-each LSTM layer focus on the important temporal details by weighting key timesteps. The last parts are dropout layers to reduce overfitting, and edge-optimized quantization and pruning to make the model lightweight and portable on-edge devices. The output layer, which is a thick layer with a sigmoid activation function, predicts odds for binary category (normal vs attack). Each of the above is carried out for each stage of the curriculum training. 

At every stage of the learning, the LIME predictions that are made on the fraction of the staged data and the features of lesser relevance or feature significance less than -0.01 are removed, and hence the model is fine-tuned by removing lower importance features and retraining, optimizing the model's learning process and improving accuracy. The iterative and staged process guarantees the robustness, flexibility, and high performance of the model in a variety of IoT attack cases. 

Finally, an ensemble model is created by stacking classifiers, including the neural network, Random Forest, and XGBoost with the above, staged learning model as the base model. This ensemble approach ensures improved generalization and performance. The model is evaluated using precision, recall, and accuracy metrics on the test dataset. The finally trained model uses LIME models to explain predictions so they can be more easily interpreted and to help the model be more transparent and trustworthy in the world. This entire workflow shows curriculum learning in conjunction with traditional datasets as progressive and adaptive learning.

\subsection{Dataset}

The proposed framework is implemented and tested on three datasets— Edge-IIoT, CIC-APT-IIoT-2024, and CIC-IoV-2024 —to evaluate its performance and robustness. A detailed description of each dataset is provided below.

The CIC-IoV-2024 dataset~\cite{carlosciciov2024} simulates cybersecurity vulnerabilities in the IoV, spoofing, and DoS attacks on a 2019 Ford car’s CAN-BUS protocol. The benchmark data reveals the insider talk, which allows scientists to explore cutting-edge security for IoV platforms. The dataset is used for testing machine learning algorithms to detect, avert, and mitigate cyber-attacks in IoVs. Specifying intra-vehicular cybersecurity, the CIC-IoV-2024 dataset represents the frontier of IoV security research with increased feature coverage and integration with other smart city networks, which will be the next frontier in vehicle security.

The Edge-IIoT dataset~\cite{9751703} provides a realistic cybersecurity data set for IoT and IIoT use cases. This dataset has been built for Machine learning-based intrusion detection systems with centralized and federated learning and is structured in seven layers Cloud Computing, Fog Computing, Edge Computing, etc. It leverages the best of the technologies such as the ThingsBoard IoT platform, OPNFV platform, and Hyperledger Sawtooth while modeling various IoT devices like temperature sensors, pH meters, and heart rate monitors. The report reports 14 attacks that are divided into 5 threats such as DoS/DDoS, Man-in-the-Middle, and Injection attacks. Having subtracted and pruned 61 features from more than 1,000 raw features, this dataset serves as a large-scale test set for machine learning techniques in IoT security.

The CIC-APT-IIoT-2024 dataset~\cite{ghiasvand2024cicaptiiotprovenancebasedaptattack}, an IIoT Advanced Persistent Threats (APT) detection dataset. It was created using a mixed testbed with real and fake IIoT parts and it is a model of attacks based on APT29 groups. More than 20 attack methods from eight primary methods, mapped to APT attack patterns are present in the dataset. The primary data types it has are provenance data in CSV (representing nodes and edges of a graph) and network logs in pcap (can be processed for feature extraction). The dataset also contains attack details including timestamps and attack types which is useful to build provenance-based detection models and test machine learning models for APT detection.

\section{Evaluation and Analysis of Proposed Model} \label{sec4}

The evaluation metrics in Table \ref{tab:evaluation_metrics} are used to show the performance of the proposed model for three datasets: Edge-IIoT, CIC-APT-IIoT-2024, and CIC-IoV-2024. The Edge-IIoT dataset had the framework score of 99\% precision, 96\% recall, and F1-Score of 97\% with an overall accuracy of 97\%, which is to say, it can accurately identify and classify both normal and attack conditions. On the CIC-APT-IIoT-2024 dataset, a more complicated attack set derived from APTs, the model performed quite well with a precision of 99\%, recall of 95\%, F1-Score of 96\%, and accuracy of 98\% which is proof of its capability to identify sophisticated and persistent threats. Similarly, on the CIC-IoV-2024 dataset (spoofing and DoS attacks in IoV environment), it got impressive results of perfect precision 100\%, recall 97\%, F1-Score 99\%, and accuracy 98\%, indicating flexibility and accuracy to secure vehicular communication networks. Not only are these findings supporting the reliability and scalability of the outlined curriculum learning model, but also demonstrate it as a robust and performant solution to diverse IoT and IIoT security problems, enabling its implementation in the field.

\begin{table}[h!]
    \centering
    \caption{Performance Metrics Across Different Datasets}
    \begin{tabular}{lcccc}
        \toprule
        \textbf{Dataset} & \textbf{Precision} & \textbf{Recall} & \textbf{F1-Score} & \textbf{Accuracy} \\
        \midrule
        1. Edge-IIoT        & 99\% & 96\% & 97\% & 97\% \\
        2. CIC-APT-IIoT-2024 & 99\% & 95\% & 96\% & 98\% \\
        3. CIC-IoV-2024      & 100\% & 97\% & 97\% & 98\% \\
        \bottomrule
    \end{tabular}
    \label{tab:evaluation_metrics}
\end{table}

\begin{figure}
    \centering
    \includegraphics[width=1\linewidth]{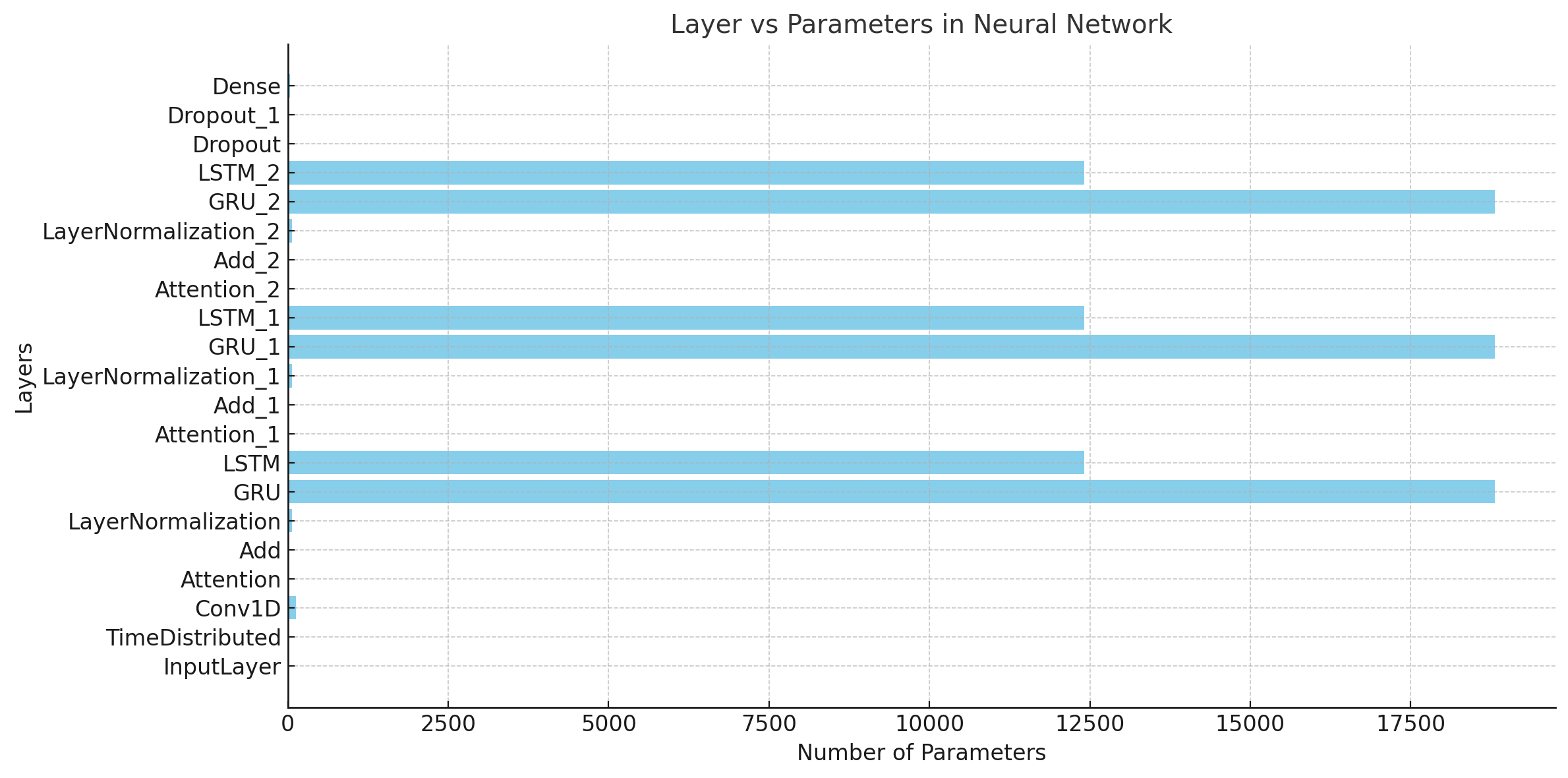}
    \caption{Layer Vs Parameters In Neural Network}
    \label{fig:layerwiseparameters}
\end{figure}

The proposed neural network model for each layer for the proposed model comprises a total of 94,051 parameters, amounting to a compact model size of approximately 367.39 KB. The breakdown of parameters across different layers, as shown in the figure~\ref{fig:layerwiseparameters}, highlights the distribution of computational complexity among the layers. Key contributors to the parameter count include GRU, LSTM, and Attention layers, which are essential for capturing temporal dependencies and enhancing model accuracy. Despite utilizing advanced components like residual connections, normalization layers, and self-attention mechanisms, the model remains lightweight, making it highly feasible for deployment on resource-constrained edge devices. The model achieves this efficiency through techniques like pruning, adaptive feature selection, and the use of GRU layers, which require fewer parameters compared to traditional LSTM layers for short-term dependencies. This lightweight design ensures scalability and real-time applicability, while maintaining high performance, making it a robust yet efficient solution for IoT and IIoT security challenges.

\subsection*{Ablation Study}

The following section presents the ablation study conducted on the proposed curriculum learning framework, evaluated using the CIC-IoV-2024 dataset.

\begin{figure}
    \centering
    \includegraphics[width=1\linewidth]{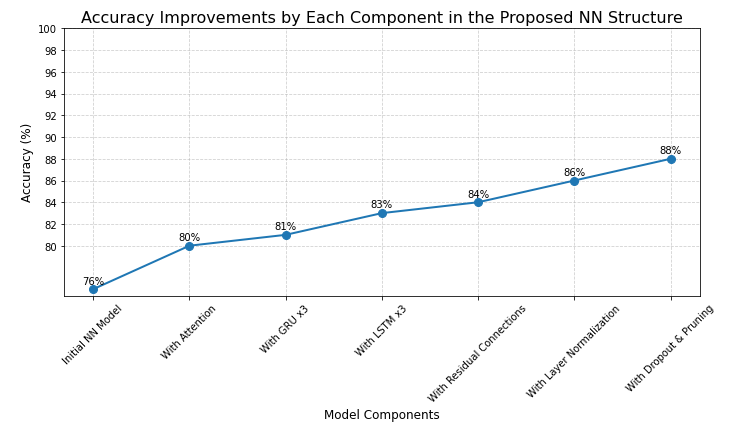}
    \caption{Accuracy Improvements by Integrating Key Components in the Proposed Neural Network Structure}
    \label{fig:accimpnn}
\end{figure}

The figure~\ref{fig:accimpnn} shows the incremental improvement in accuracy by including certain pieces of the NN design. If the baseline model had an accuracy of 76\%, introducing the attention techniques improves the accuracy to 80\% showing how to pay attention to relevant temporal parameters. Add three GRU layers to the model and the performance is further improved to 81\%, due to their effectiveness in short-term dependency detection. With 3 LSTM layers, accuracy is improved to 83\%, and they can teach long-term dependencies. - Remainder connections facilitate better gradient encoding to increase accuracy up to 84\%. Layer normalization also smooths training to 86. Lastly, edge-optimization algorithms such as dropout and pruning reduce over-fitting which results in 88\% accuracy. These enhancements prove how efficient each piece is to generate a reliable and performant proposed NN model for IoT network security scenarios.

\begin{table*}[h!]
    \centering
    \caption{Ablation Study: Accuracy Improvements}
    \begin{tabular}{lccl}
        \toprule
        \textbf{Model Component} & \textbf{Improvement (\%)} & \textbf{Cumulative Accuracy (\%)} \\
        \midrule
        Base Neural Network                 & -   & 88  \\
        With Curriculum Learning                & +6  & 94  \\
        With XAI (LIME) Integration and Un-Learning & +3  & 97  \\
        Stacking Ensemble                       & +1  & 98  \\
        \bottomrule
    \end{tabular}
    \label{tab:ablation_study}
\end{table*}

The table~\ref{tab:ablation_study} above is an ablation study of the incremental increase in model precision due to encoding some framework elements. Starting from the baseline neural network, the accuracy is 88\%. Curriculum learning is also included, increasing the accuracy by 6\% and showing how staged training can be effective in gradually improving the model’s performance in tackling difficult data structures. Explainable AI (XAI) with LIME and feature un-learning gives 3\% additional improvement as the model learns to learn more relevant features and discard the ones that aren’t important. Lastly, if we use a stacking ensemble, it is 1\% more accurate, and the total accuracy is 98\%. Such findings illustrate the need for a robust and generalizing approach through curriculum learning, explainable AI, and ensemble strategies combined.

\begin{figure}
    \centering
    \includegraphics[width=1\linewidth]{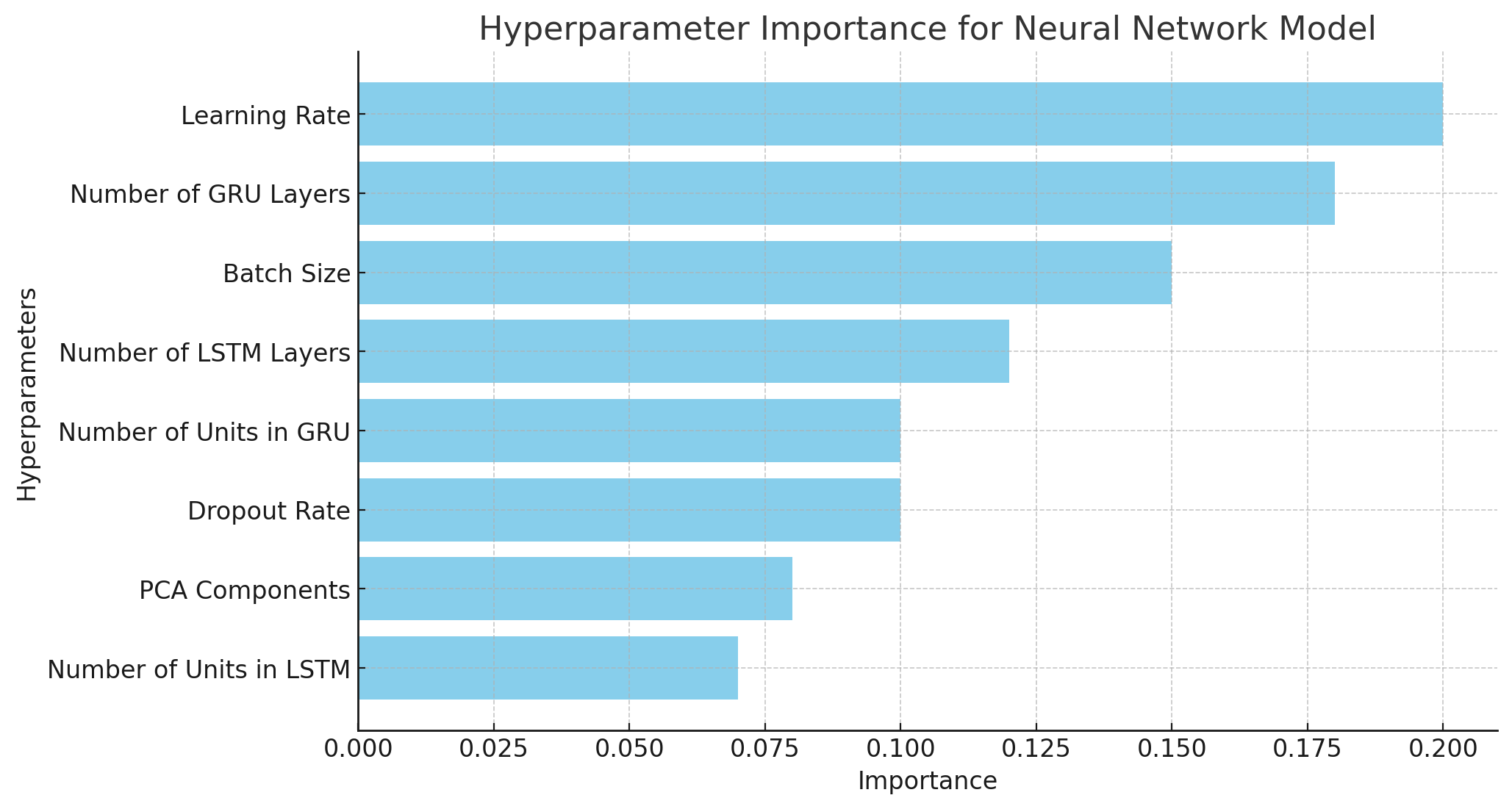}
    \caption{Hyper-Parameters Importance in the Proposed Neural Network Structure}
    \label{fig:hyperparameters}
\end{figure}

The fig.~\ref{fig:hyperparameters} illustrates the importance of various hyperparameters in influencing the performance of the proposed neural network model. Key hyperparameters like the learning rate, number of GRU layers, and batch size exhibit the highest impact, reflecting their critical roles in optimizing the model's convergence and capacity. The number of LSTM layers and GRU units also significantly contributes to capturing temporal dependencies in the sequential data. Meanwhile, dropout rate, PCA components, and the number of LSTM units demonstrate moderate importance, indicating their influence on reducing overfitting, dimensionality reduction, and feature extraction. These insights underline the necessity of careful tuning of hyperparameters to achieve an optimal trade-off between model complexity and performance. Prioritizing high-impact parameters during optimization can streamline model refinement and ensure efficient resource utilization.

\begin{figure*}
    \centering
    \fbox{\includegraphics[width=1\linewidth]{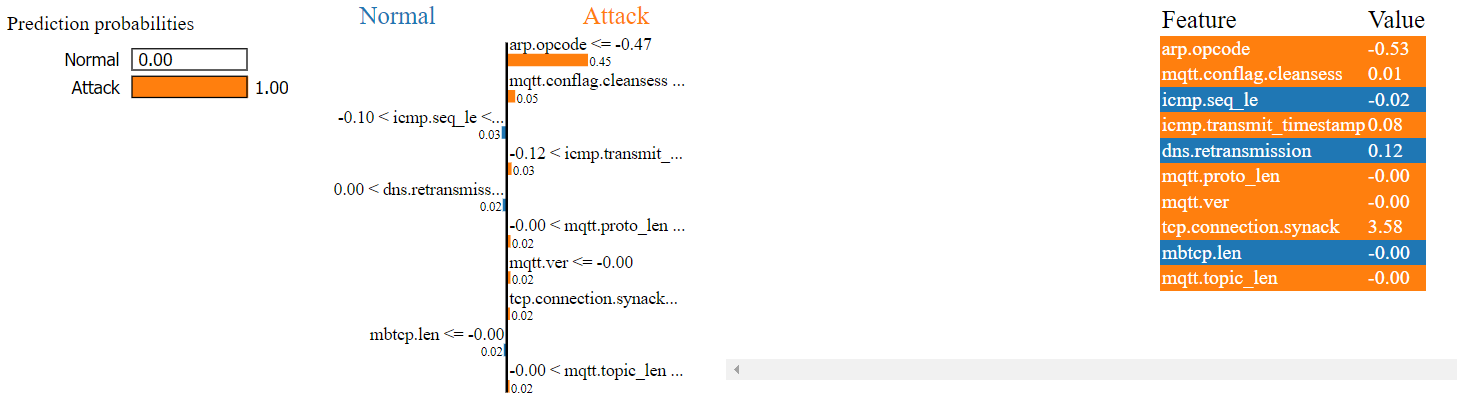}}
    \caption{Example LIME Explanation for Attack Classification, Highlighting Key Features and Their Contributions to the Model's Prediction}
    \label{limepred}
\end{figure*}

The figure.~\ref{limepred} represents an example of a LIME (Local Interpretable Model-Agnostic Explanations) explanation for a data point classified as an attack. These LIME explanations are integrated into the proposed model to enhance interpretability and facilitate informed decision-making. On the left, the prediction probabilities highlight that the model confidently identifies the data as an attack with a probability of 1.0. The center section provides a breakdown of the contributing features, with their thresholds and respective weights toward the attack classification. On the right, the feature values are listed, showing how specific attributes, such as 'arp.opcode' and 'dns.retransmission', strongly influence the decision-making process. This detailed insight enables researchers to understand the key factors contributing to the model's predictions, enhancing interpretability and trust in the decision-making process for detecting attack patterns.

The results of the model shown here prove it’s effective in terms of accuracy, simplicity of design, and powerful training through curriculum learning. It proved very good for various types of datasets, 97\% on Edge-IIoT, 98\% on CIC-APT-IIoT-2024, and 98\% on CIC-IoV-2024, with various attack complexity and IoT/IIoT security issues handled. The lightweight implementation (94,051 parameters (367.39 KB) guarantees scalability and real-time use on resource-limited edge devices). This efficiency is achieved by pruning, adaptive feature selection, and the intelligent use of GRU/LSTM layers for short-term and long-term dependency capture respectively. Curriculum learning also contributed to enhancing the power of model training by slowly testing the model with more and more attack scenarios, so that it could become generalizable and perform better. XAI (LIME) integration also improved interpretability and feature fit, and stacking ensembles provided small but useful accuracy improvements. Together, these results affirm the model's potential as a reliable, efficient, and interpretable solution for IoT and IIoT network security.

\section{Conclusion} \label{sec5}

In conclusion, the proposed curriculum learning framework demonstrated exceptional performance and robustness across diverse IoT and IIoT datasets, including Edge-IIoT, CIC-APT-IIoT-2024, and CIC-IoV-2024. The model achieved high accuracy rates of 97\%, 98\%, and 98\% respectively, with precision, recall, and F1-scores exceeding 95\% in all cases, showcasing its ability to accurately classify normal and attack conditions. The lightweight neural network, with 94,051 parameters (367.39 KB), ensures scalability and feasibility for edge device deployment. The integration of curriculum learning significantly improved model training by progressively tackling data complexity, while XAI techniques like LIME enhanced interpretability and feature refinement. These results underline the framework’s potential as a robust, efficient, and interpretable solution for addressing complex IoT security challenges.

\section{Future Scope} \label{sec6}

The future direction of this research would be to expand the proposed curriculum learning framework to cover changing and dynamic attack trends in real-time IoT scenarios. Federated learning methods can add to data privacy and security by providing edge-device distributed training. Moreover, by covering larger datasets with different IoT ecosystems (healthcare, smart cities, etc.), we can increase the flexibility and generalizability of the framework. Combining sophisticated XAIs (e.g., SHAP or counterfactual explanations) with model decisions gives you better insights into model choices and makes them more trustworthy and intuitive. Lastly, hardware acceleration approaches such as quantization-based training or run-on custom edge AI hardware can also further enable scaling the system in power and limited resources applications to industrial and consumer IoT.


\appendix

\end{document}